\documentclass[aps,prb,twocolumn,amsmath,amssymb,showpacs]{revtex4}
\usepackage{graphicx}
\usepackage{dcolumn}
\usepackage{bm}
\usepackage{color}
\usepackage{here}
\graphicspath{{fig/}}
\begin{document}

\title{Dynamics of the quantum dimer model on the triangular lattice: Soft modes and 
local resonating valence-bond correlations.}

\author{Arnaud Ralko,${^1}$ Michel Ferrero,${^2}$ Federico Becca,${^{1,2}}$ Dmitri Ivanov,${^1}$ 
and Fr\'{e}d\'{e}ric Mila${^1}$}
\affiliation{
${^1}$ Institute of Theoretical Physics, Ecole Polytechnique F\'{e}d\'{e}rale de Lausanne 
(EPFL), CH-1015 Lausanne, Switzerland \\
${^2}$ INFM-Democritos, National Simulation Center and International School for Advanced 
Studies (SISSA), I-34014 Trieste, Italy}

\date{July 24, 2006}

\begin{abstract}
We report on an exhaustive investigation of the dynamical dimer-dimer correlations 
in imaginary time for the quantum dimer model on the triangular lattice using the Green's 
function Monte Carlo method. We show in particular that soft modes develop upon reducing 
the dimer-dimer repulsion, indicating the presence of a second-order phase transition
into an ordered phase with broken translational symmetry. We further investigate the nature
of this ordered phase, for which a 12-site unit cell has been previously proposed, with the 
surprising result that significant Bragg peaks are only present at two of the 
three high-symmetry points consistent with this unit cell. We attribute the absence
of a detectable peak to its small magnitude due to the nearly uniform
internal structure of the 12-site crystal cell.
\end{abstract}

\maketitle

\section{Introduction}

Resonating valence-bond (RVB) states constitute a major theme in strongly correlated
systems, both in the context of Mott insulators and of 
superconductors.~\cite{anderson,misguich} 
In his milestone paper, Anderson proposed that high-temperature 
superconductors are intimately related to Mott insulators and that the pairing mechanism 
is due to ``preformed'' pairs already present in the strongly correlated insulating 
phase.~\cite{anderson} Originally, the RVB phase was defined by a fully-projected
BCS wave function where the electron pairs have an arbitrary range. 
A simplification of the RVB construction was further proposed by Rokhsar and Kivelson,
who considered an effective quantum dimer model (QDM) with only local processes and
orthogonal dimer coverings.~\cite{rokhsar}
The conditions under which such a Hamiltonian could be an accurate description
of an SU(2) Heisenberg model are not fully understood yet, but it is expected to 
be a reasonable approximation if the dimer coverings constitute a natural variational 
basis, and specific quantum dimer models have recently been derived for the trimerized 
Kagome antiferromagnet~\cite{mila} by Zhitomirsky,~\cite{zhitomirsky} and for a 
spin-orbital model by Vernay and collaborators.~\cite{vernay}

Regardless of its actual validity for microscopic spin models, the QDM of Rokhsar and
Kivelson has attracted a lot of attention as a promising way to investigate RVB physics. 
In particular, the QDM on the triangular lattice has been shown to possess
a liquid phase with an exponential decay of all correlations.~\cite{moessner}
The Hamiltonian of that model is
\begin{eqnarray}\label{hamilt}
H &=& -t \sum
\left(
|\unitlength=1mm
\begin{picture}(6.2,5)
\linethickness{0.5mm}
\put(0.9,-.7){\line(1,2){1.8}}
\put(3.8,-.7){\line(1,2){1.8}}
\end{picture}
\rangle
\langle
\unitlength=1mm
\begin{picture}(6.5,5)
\linethickness{0.3mm}
\put(3.2,2.6){\line(1,0){3.2}}
\put(0.9,-.7){\line(1,0){3.2}}
\end{picture}
|
+h.c.\right) \nonumber \\ 
&+& \ V \sum \left(
|\unitlength=1mm
\begin{picture}(6.2,5)
\linethickness{0.5mm}
\put(0.9,-.7){\line(1,2){1.8}}
\put(3.8,-.7){\line(1,2){1.8}}
\end{picture}
\rangle
\langle
\unitlength=1mm
\begin{picture}(6.2,5)
\linethickness{0.5mm}
\put(0.9,-.7){\line(1,2){1.8}}
\put(3.8,-.7){\line(1,2){1.8}}
\end{picture}|+
|
\unitlength=1mm
\begin{picture}(6.5,5)
\linethickness{0.3mm}
\put(3.2,2.6){\line(1,0){3.2}}
\put(0.9,-.7){\line(1,0){3.2}}
\end{picture}\rangle
\langle
\begin{picture}(6.5,5)
\linethickness{0.3mm}
\put(3.2,2.6){\line(1,0){3.2}}
\put(0.9,-.7){\line(1,0){3.2}}
\end{picture}
|
\right),
\end{eqnarray}
where the sum runs over all plaquettes including the three possible orientations.
The kinetic term controlled by the amplitude $t$ changes the dimer covering of every 
flippable plaquette, i.e., of every plaquette containing two dimers facing each other, 
while the potential term controlled by the interaction $V$ describes a repulsion ($V>0$) 
or an attraction ($V<0$) between dimers facing each other.

Using the Green's function Monte Carlo (GFMC) algorithm to probe the ground-state properties
of large clusters, we recently proved the existence, in the thermodynamic limit, 
of topological degeneracy in the disordered (i.e., RVB) phase.~\cite{ralko} This degeneracy persists 
over an extended parameter range below the Rokhsar-Kivelson point $V/t=1$. With decreasing
$V/t$, the RVB phase is replaced by a crystalline phase with a 12-site unit cell, the so-called
$\sqrt{12} \times \sqrt{12}$ phase. Upon further decreasing the dimer-dimer repulsion,
this phase is ultimately followed by a columnar phase for sufficiently large and negative $V/t$.

From the point of view of RVB physics, one of the most interesting questions is
the nature of the crystallization transition
between the RVB and the $\sqrt{12} \times \sqrt{12}$ phases. In the present work,
we analyze this phase transition by extending the GFMC algorithm to calculate 
dynamical dimer-dimer correlations. This has allowed us to extract the energy of the first 
excited state throughout the Brillouin zone (BZ). Having in mind
that the $\sqrt{12} \times \sqrt{12}$ phase has dimer order in the thermodynamic limit, 
the dimer gap must close at some k-points of the BZ upon entering that phase. 
An analysis of the gap behavior (on the disordered side)
and of the structure factor (on the ordered side) near the transition point gives
a strong evidence in favor of the second order phase transition.

The paper is organized as follows: In Section~\ref{method}, we give a brief
explanation of how we calculate the dynamical correlations and discuss the 
importance of the choice of the finite-size clusters. Then, in Section~\ref{result1} 
we discuss the excitation spectrum, as well as the location and the nature of the phase 
transition between the RVB liquid and the crystalline phase. Finally, Section~\ref{result2}
is dedicated to the internal structure of the $\sqrt{12} \times \sqrt{12}$ phase. 
Concluding remarks are given in Section~\ref{conc}. 

\section{The numerical method}\label{method}

The numerical investigations are based on a zero-temperature Monte Carlo method that
filters out the high-energy components of a starting wave function $|\Psi_G \rangle$ by
the iterative application of $G=(\Lambda-H)$, where $\Lambda$ is a diagonal 
matrix proportional to the identity. This allows us to sample directly the ground state 
$|\Psi_0 \rangle$. Indeed, in the case of the QDM defined by 
Eq.~(\ref{hamilt}), all the off-diagonal matrix elements are non-positive and, therefore, 
there is no sign problem and the GFMC is numerically exact.~\cite{nandini}
Moreover, besides the ground-state energy and static correlation functions, 
it is also possible within the same method to compute dynamical correlations in imaginary 
time, such as the dimer-dimer correlation function defined by
\begin{equation}\label{dimer1}
D(k,\tau) = \frac{\langle \Psi_G| d(-k) e^{-H \tau} d(k) |\Psi_0 \rangle}
{\langle \Psi_G| e^{-H \tau} |\Psi_0 \rangle},
\end{equation}
where $d(k)$ is the Fourier transform of the dimer operator $d(r)$ in real space 
that is equal to $1$ if there is a dimer at site $r$ in the $(1,0)$ direction and $0$ 
otherwise. The expression of Eq.~(\ref{dimer1}) can be easily computed by GFMC by applying 
the so-called forward-walking technique.~\cite{calandra} This method to calculate dynamic 
correlation functions has been recently used to investigate an $XY$ spin model with 
ring-exchange.~\cite{spanu} 
In practice, after an equilibration time, the dimer configurations $|x \rangle$ are 
statistically sampled. Then the (diagonal) dimer operator $d(k)$ is applied to $|x \rangle$
and the system is evolved for $L$ further steps by the forward walking.
Within this technique, it is very important to {\it analytically} perform the limit of
$\Lambda \to \infty$,~\cite{caprio} which allows us to speed up the algorithm and define
the imaginary time $\tau$. In more detail, in the limit $\Lambda \to \infty$, the 
probability of the diagonal moves grows and, therefore, it is better to sample directly 
the number of diagonal moves before changing the configuration $|x \rangle$. 
Note that, in order to have a finite probability for non-diagonal moves 
for $\Lambda \to \infty$, we have to apply $G$ to a power proportional to $\Lambda$, 
i.e., $G^{\beta \Lambda}$. Then
\begin{equation}
\lim_{\Lambda \to \infty} \left ( \Lambda - H \right )^{\beta \Lambda} \sim e^{-H \beta}.
\end{equation}
This algorithm allows us to perform at once all the diagonal moves and then perform
directly the non-diagonal one, thus saving a huge computational effort. Then
\begin{equation}
D(k,L\beta) = \lim_{M \to \infty} 
\frac{\langle \Psi_G| d(-k) e^{-H L\beta} d(k) e^{-H M\beta} |\Psi_G \rangle}
{\langle \Psi_G| e^{-H (L+M)\beta} |\Psi_G \rangle},
\end{equation}
and the imaginary time of Eq.~(\ref{dimer1}) is given by $\tau=L\beta$.
It is easy to demonstrate that for large $\tau$, when all the high-energy states have been
projected out, we have 
\begin{equation}
D(k,\tau) \propto e^{- \Delta(k) \tau}.
\end{equation}
Then, by fitting the large-$\tau$ behavior of $D(k,\tau)$, it is possible to 
estimate the gap $\Delta(k)$ of the first excited state at the k-point in the BZ.
Note that, within this approach, we only find the excitations that have a {\it finite} 
overlap with the state $d(k) |\Psi_0 \rangle$. In this work, we are only interested in 
the lowest excitation energy, which can be extracted by considering only one exponential 
and fitting the largest-$\tau$ regime. In order to verify the accuracy on $\Delta(k)$, 
we have also checked the results by fitting with two and three exponentials. For extracting
more information on the spectrum, one may implement more involved methods 
such as the maximum-entropy technique,~\cite{gubernatis} which goes beyond the
scope of the present paper.

\begin{figure}
\includegraphics[width=0.4\textwidth]{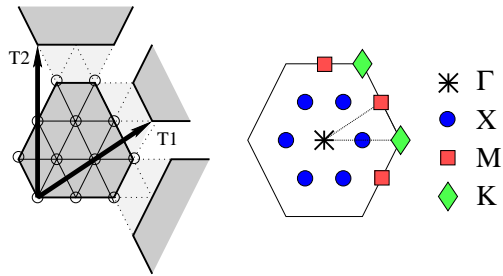}
\caption{\label{fig:cluster} 
(Color online) Left: Example of cluster used in the calculations (here the 12-site one). 
Right: Brillouin zone of the triangular lattice. High-symmetry points are labelled 
by $K=(4\pi/3,0)$, $M=(\pi,\pi/\sqrt{3})$, and $X=(2\pi/3,0)$.}
\end{figure}

Since we are interested in the transition between the disordered RVB phase 
and the crystalline $\sqrt{12} \times \sqrt{12}$ one, we have decided to work only with 
clusters defined by factorization of the infinite triangular lattice by the
translations~\cite{bernu}
\begin{eqnarray}
{\bf{T}}_{1} &=& l {\bf{u}}_{1} + l {\bf{u}}_{2}  \nonumber \\
{\bf{T}}_{2} &=& -l {\bf{u}}_{1} + 2l {\bf{u}}_{2} \nonumber ,
\end{eqnarray}
where ${\bf u}_1=(1,0)$ and ${\bf u}_2=(1/2,\sqrt{3}/2)$ are the unitary vectors defining 
the triangular lattice and $l$ is an integer. Such clusters can accommodate the
proposed crystal structure without defects. The number of sites in the
cluster is $N=3 l^2$, and to admit dimer coverings $l$ must be even.
An example of such a cluster for $l=2$ is shown in Fig.~\ref{fig:cluster}.
The right panel of  Fig.~\ref{fig:cluster} shows the BZ of the triangular lattice 
with the high-symmetry points $K=(4\pi/3,0)$, $M=(\pi,\pi/\sqrt{3})$, and $X=(2\pi/3,0)$.

We remind that the QDM on a toroidal cluster has four topological sectors denoted
by $(0,0)$, $(1,0)$, $(0,1)$, and $(1,1)$: The Hilbert space splits into four
subspaces that cannot be connected by local dimer flips like the ones contained in the
Hamiltonian. Given the rotational symmetry of the cluster, three of the four topological
sectors are always exactly degenerate, and the only independent ones are the $(0,0)$ and the $(1,1)$,
in the notation of Ref.~\onlinecite{ralko}.
The ground state belongs to the former or the latter depending on the
parity of $l/2$.~\cite{ralko}
It is worthwhile to mention that, within GFMC, we can fix the topological sector during the
simulation and, therefore, we can study the different sectors independently.

In the following, we will present the results for the dimer gap along particular symmetry
directions of the BZ, and in particular the size scaling of the k-points where
soft modes are expected to develop upon entering the crystalline 
$\sqrt{12} \times \sqrt{12}$ phase.

\section{The RVB -- crystal transition}\label{result1}

In their original work, Moessner and Sondhi~\cite{moessner} estimated that the transition 
from the disordered to a crystalline phase occurred around $V/t \simeq 0.7$.
A later analysis of the topological splitting on finite-size clusters suggested
a slightly larger value ($0.7 < V/t < 0.85$).~\cite{ralko}
The analysis of dynamic correlations reported in our present work provides a
more accurate estimate, $V/t=0.82\pm 0.03$, as will be demonstrated below.

In the crystalline phase, the existence of a broken translational symmetry naturally 
implies a folding of the BZ and soft modes at particular k-points. In particular, 
for the 12-site unit cell proposed by Moessner and Sondhi, we expect to find soft modes 
at the $X$, $M$, and $K$ points (see Fig.~\ref{fig:cluster}). 
So we now turn to an investigation of the excitation spectrum as revealed by our GFMC 
approach, with emphasis on the behavior around these high symmetry points.

\begin{figure}
\includegraphics[width=0.4\textwidth]{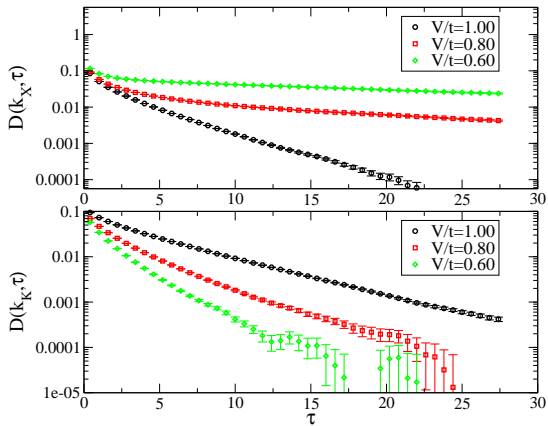}
\caption{\label{fig:dyncorr} 
(Color online) Typical results for the dynamical dimer-dimer correlations $D(k,\tau)$ for the 
432-site cluster at $X=(2\pi/3,0)$ (upper panel) and $K=(4\pi/3,0)$ (lower panel) for 
a few selected values of $V/t$.}
\end{figure}

\begin{figure}
\includegraphics[width=0.4\textwidth]{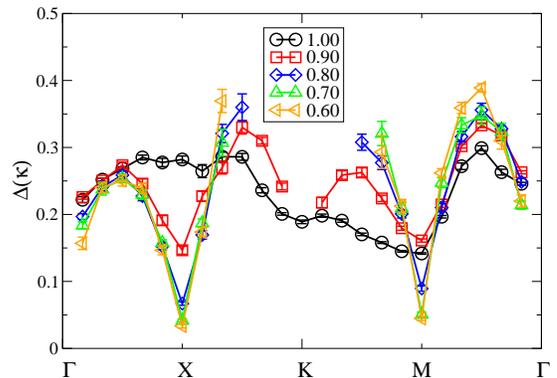}
\caption{\label{fig:spectrum} 
(Color online) Lowest-energy excitation spectra for the 432-site cluster for different values 
of $V/t$.}
\end{figure}

We perform a systematic analysis of the dynamical correlations for clusters with up to 432 
sites and for $0.55 \leq V/t \leq 1.0$. We report in Fig.~\ref{fig:dyncorr} the results 
for a 432-site cluster ($N=3 l^2$ with $l=12$) at $X$ and $K$ for
different values of $V/t$. Although in principle the long-time dynamics should always 
give access to the energy of the first excited states, in practice the error bars grow with
time, and two types of data sets may be distinguished. In the first, favorable case,
an exponential decay governed by a single exponent shows up long before the error bars
become significant, and the gap can be extracted very accurately (see upper panel
of Fig.~\ref{fig:dyncorr}). In a logarithmic scale, this shows up as a linear behavior of 
$D(k,\tau)$ prior to the development of large error bars. 
This is the case throughout the BZ at the
Rokhsar-Kivelson point ($V/t=1$), and everywhere except around the $K$ point when $V/t<1$.
In the other case, the error bars start to grow before a simple exponential decay has set in.
In logarithmic scale, this shows up as a persistent curvature. An example is reported
in the lower panel of Fig.~\ref{fig:dyncorr}. The absence of a clear exponential regime
prevents us from extracting a reliable value for the energy of the first excited state.

The results of this analysis are depicted in Fig.~\ref{fig:spectrum}, where we show the 
excitation spectrum of the 432-site cluster in the topological sector $(0,0)$ (the one 
that contains the ground state) for different values of $V/t$.
The missing points around $K$ correspond to cases where the extraction of the energy was 
impossible. This spectrum calls for several comments. 

First of all, the spectrum is fully gapped close to the
Rokhsar-Kivelson point, in agreement with the results obtained previously by Ivanov
at the Rokhsar-Kivelson point,~\cite{ivanov} and consistent 
with a liquid ground state with no long-range order. 

\begin{figure}
\includegraphics[width=0.4\textwidth]{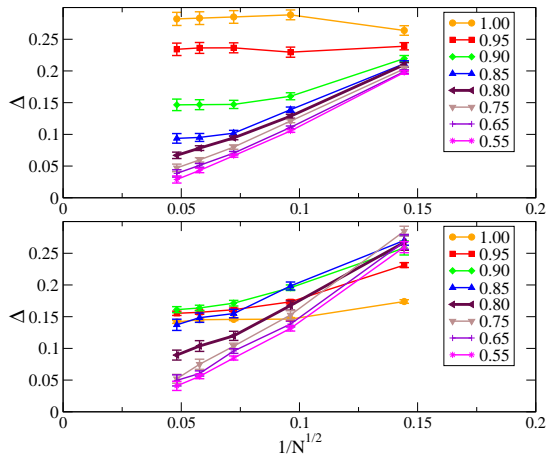}
\caption{\label{fig:sizescale} 
(Color online) Size scalings of the gap at the $X$ (upper panel) and $M$
(lower panel) points  for different values of $V/t$.}
\end{figure}

Secondly, upon decreasing $V/t$, two soft modes clearly appear at $X$ and $M$.
These two k-points are compatible with the 12-site unit cell structure of the 
$\sqrt{12} \times \sqrt{12}$ phase, proposed by Moessner and Sondhi.~\cite{moessner2}
It turns out that the numerical results concerning $X$ and $M$ are very accurate and allow 
us to obtain a rather precise determination of the transition point. 
In Fig.~\ref{fig:sizescale}, we show the size scaling of these two gaps. 
At the Rokhsar-Kivelson point, the curves saturate quickly to a finite value in the 
thermodynamic limit. This value becomes smaller by decreasing the ratio $V/t$, and
we can estimate that the transition occurs around $V/t \simeq 0.8$, not far from
our previous estimate based on the topological gap.~\cite{ralko}
This constitutes one of the important results of this paper: The transition out
of the RVB phase is second order, as revealed by the development of soft modes
upon entering the crystalline phase.

Thirdly, and quite surprisingly, we could not get any evidence for the emergence of a soft 
mode at the $K$ point, the third high-symmetry point compatible with the 12-site unit cell.
On the contrary, the slope of $D(K,\tau)$ (in logarithmic scale) at the largest 
accessible times is still very large, so that there is no evidence whatsoever of a 
small energy scale at that point. We can think of two possibilities to explain this result:
i) The overlap between the first excitation and the starting state $d(k)|\Psi_0 \rangle$ 
is so small that this excitation does not give any detectable contribution to the 
dynamics in the accessible time range; 
ii) The symmetry of the ordered phase is such that no soft mode develops at the $K$ point.
To clarify this point, we now turn to a detailed investigation of this crystalline phase.

\begin{figure}
\includegraphics[width=0.4\textwidth]{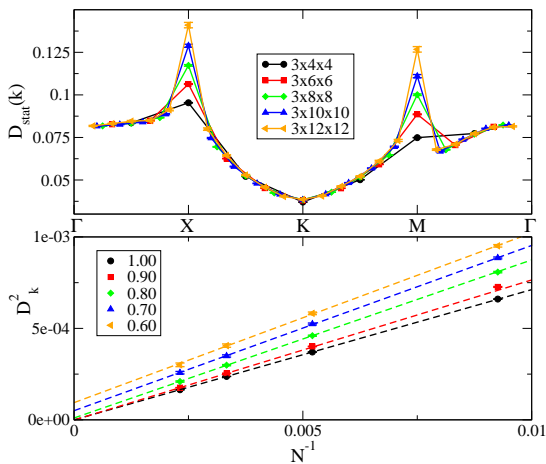}
\caption{\label{fig:statcorr} 
(Color online) Static dimer-dimer correlation at $V/t=0.55$ for different cluster sizes 
(upper panel). Size scaling of the dimer order parameter square $D_k^2$ 
(calculated at the $X$ point) for different values of $V/t$ (lower panel).} 
\end{figure}

\section{The $\sqrt{12} \times \sqrt{12}$ phase}\label{result2}

The unexpected behavior found at $K$ is actually not in contradiction with any of the
previous numerical studies of the static correlations in that phase: Moessner and Sondhi detected a 
peak at $X$ and noticed it was compatible with a 12-site unit cell, but they did not 
calculate the full k-dependence of the static dimer-dimer correlations.~\cite{moessner} 
In our previous paper,~\cite{ralko} we concentrated on the RVB phase and showed that the 
correlations were indeed decreasing very fast in that phase, but we did not investigate in 
details the correlations in the ordered phase. So the first thing to do is to go back to the 
static correlations and to calculate the full k-dependence of the static dimer-dimer 
correlation function defined by:
\begin{equation}
D_{\rm stat}(k) = \frac{\langle \Psi_0| d(-k) d(k) |\Psi_0 \rangle}{\langle \Psi_0|\Psi_0 \rangle}.
\end{equation}
The results for different cluster sizes and $V/t=0.55$ are depicted in 
Fig.~\ref{fig:statcorr}. Clearly, these results confirm the anomaly of the
$K$ point: While strong peaks are present at $X$ and $M$ and grow 
with the size of the cluster, no peak could be detected at $K$. Notice that for this
kind of static correlations, there is no fitting procedure, in contraxt to the extraction
of the gap in the dynamical calculations.

\begin{figure}
\includegraphics[width=0.4\textwidth]{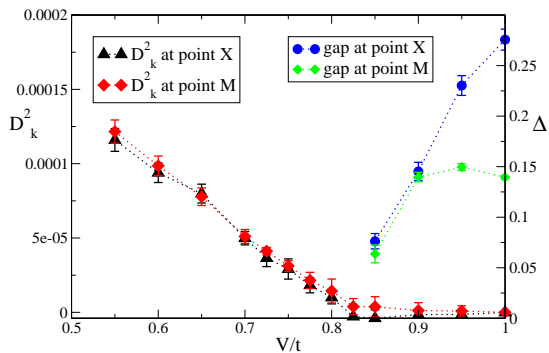}
\caption{\label{fig:order} 
(Color online) $V/t$ dependence of the thermodynamic values of $D_k^2$
and of the gap at the $X$ and $M$ points.}
\end{figure}

Before trying to understand what is going on at the $K$ point, let us look
more closely at the $X$ and $M$ points. For small enough $V/t$, the size of
these peaks grows linearly with the number of sites of the cluster, a clear
indication that Bragg peaks develop. Quantitatively, we extract the
square of the order parameter defined by:
\begin{equation}
D^2_k = \frac{D_{\rm stat}(k)}{N},
\end{equation}
for both $X$ and $M$ from the finite-size scaling of these peaks. The results
are depicted in Fig.~\ref{fig:statcorr}. The extrapolated values are very
small even well inside the crystalline phase, but given the accuracy of the data and the
smoothness of the size scaling, these values are reliable, and indeed
they again point to a second-order phase transition located between $V/t \simeq 0.8$ and
$0.85$  where the order
parameters continuously vanish at both points (see Fig.~\ref{fig:order}). So the existence 
of a crystalline phase with Bragg peaks at $X$ and $M$ is well established.

A similar analysis of the Bragg peak at the $K$ point gives a zero value
for $D^2_k$, within
the error bars. For example, well inside the crystal phase at $V/t=0.55$ we can estimate
an upper bound $D^2_k < 1.5 \times 10^{-6}$.

As mentioned at the end of the previous section, there are two possible 
explanations for the absence of the $K$ peak. The first possibility is an
exact cancellation of the $K$ peak due to the symmetry of the 
crystal phase. The second one is just a strong suppression of the
$K$ peak so that its magnitude is below our numerical precision.

\begin{figure}
\includegraphics[width=0.47\textwidth]{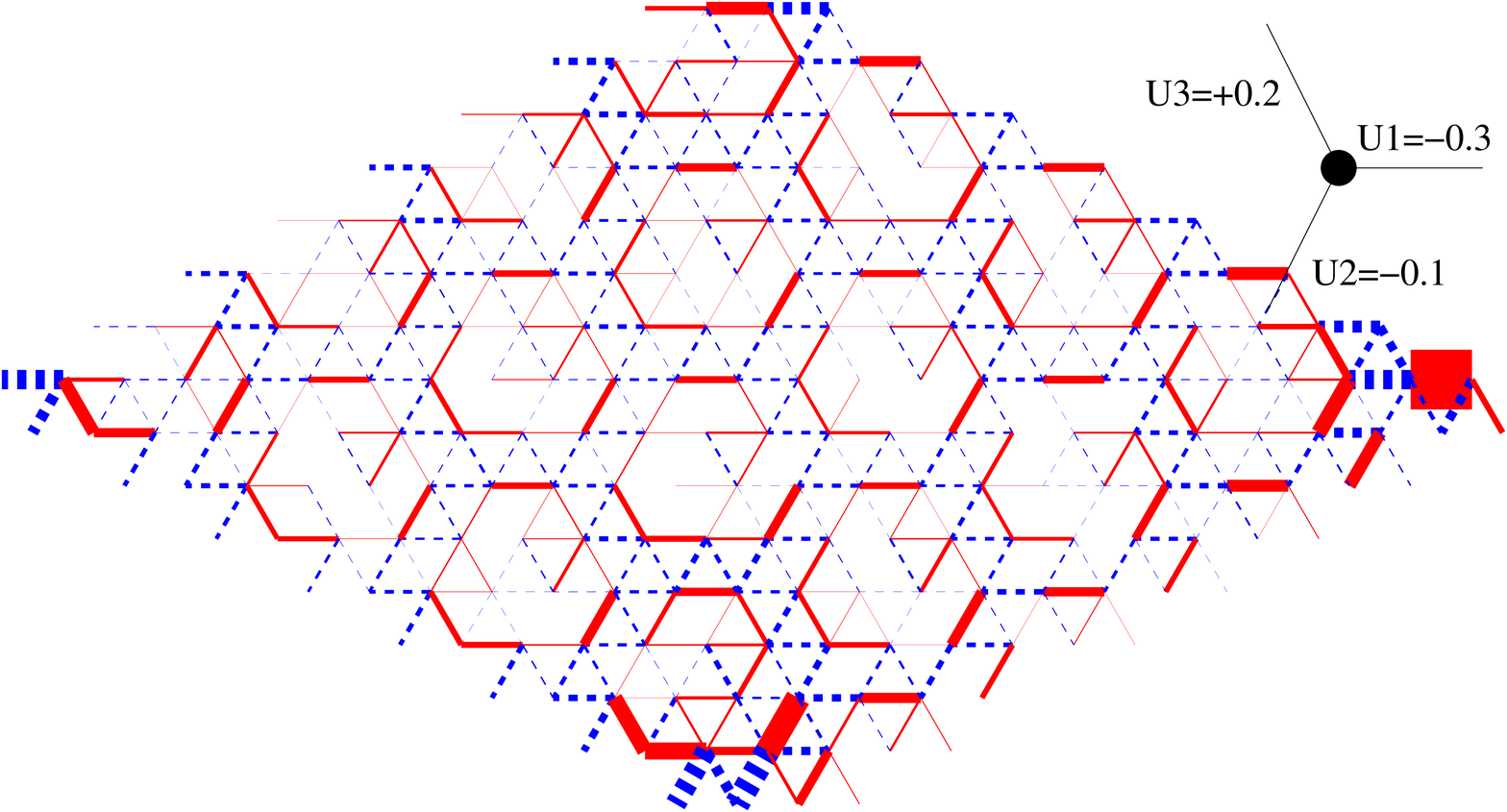}
\caption{\label{fig:density} 
(Color online) Mean value of the dimer operator for the 192-site cluster for $V/t=0.6$ in 
the presence of the local perturbation (shown in the inset) placed at the rightest site.
Dashed lines correspond to correlations lower than the uniform distribution (i.e., $1/6$)
and continuous lines to correlations larger than $1/6$. The thickness of the bonds indicates 
how  different the amplitude is from the uniform distribution.}
\end{figure}

The symmetry cancellation of the $K$ peak might occur in a crystal
phase with a three-fold rotational symmetry about a {\it lattice
site}.~\cite{three-fold-symmetry} However this type of crystal symmetry
seems to be unlikely, since it cannot accommodate any individual dimer
or plaquette covering with the symmetry of the crystal (due to the
single site in the center of the three-fold symmetry). Note that
the $\sqrt{12} \times \sqrt{12}$ crystal structure proposed by
Moessner and Sondhi~\cite{moessner2} has a three-fold symmetry
about a {\it triangular plaquette center} and should not lead
to an exact cancellation of the $K$ peak. To get a further
insight in the symmetry of the $\sqrt{12} \times \sqrt{12}$ phase,
we have calculated the static density profile in the presence
of a local perturbation breaking all symmetries of the lattice.
In this way, instead of obtaining a uniform linear combination 
of all symmetry related crystal states, the system selects one (or
sometimes several) crystal state(s) favored by the perturbation, giving access to 
a real-space snapshot of the broken-symmetry ground state
far enough from the perturbation. 
A typical result obtained by considering local potentials
for three bonds related by a $2 \pi/3$ rotation around a site that favor or
disfavor the presence of a dimer is depicted in Fig.~\ref{fig:density}.
The 12-site diamond of the type proposed by Moessner and Sondhi~\cite{moessner2}
is clearly visible. Other choices of potential perturbation sometimes
produce less symmetric periodic patterns which we interpret as superpositions
of several crystal states. However, in no pattern were we able to detect
a three-fold symmetry about a lattice site. Therefore we conclude that
the symmetry-cancellation scenario is not realized and that we should
interpret the absence of a visible $K$ peak as a strong suppression of the
corresponding Fourier harmonic in the  $\sqrt{12} \times \sqrt{12}$ state
of Moessner and Sondhi.~\cite{moessner2}

\begin{figure}
\includegraphics[width=0.4\textwidth]{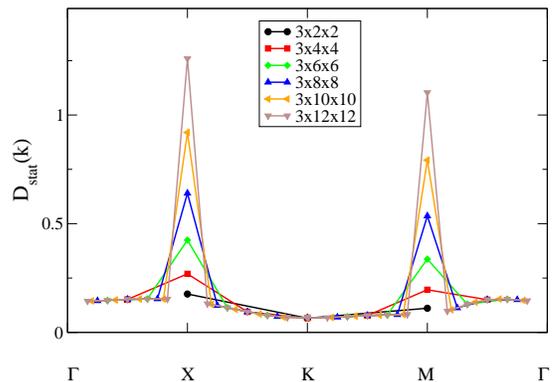}
\caption{\label{fig:trial} 
(Color online) Static dimer correlations in the trial wave function
constructed as the product of uniform RVB superpositions of dimer covering
of 12-site hexagons.}
\end{figure}

Interestingly enough, while the unit cell is clearly visible in Fig.~\ref{fig:density}, 
there is no internal modulation of the dimer density inside the resonating
12-site hexagons. Motivated by these observations, we have performed a calculation with a 
trial wave function containing a uniform RVB superposition of all dimer configurations within 
12-site hexagons and no dimers between the hexagons. 
The resulting peak weights $D_X^2$ and $D_M^2$ are of the same order of magnitude, while the 
peak weight $D_K^2$ is more than two orders of magnitude smaller. Such a huge difference 
between the peak weights would explain our numerical findings about the absence
of the $K$ peak for both dynamic and static correlations.
For illustration, the static dimer-dimer correlations for our trial wave function
are shown in Fig.~\ref{fig:trial}: They qualitatively resemble the actual
correlations in the crystal phase of Fig.~\ref{fig:statcorr}, with no visible
$K$ peak.
\section{Discussion}\label{conc}

To summarize, we have studied the phase transition between the RVB and the crystal
states of the QDM model on the triangular lattice by analyzing the dimer excitation
spectrum with the help of the Green's function Monte Carlo method.
First of all, we have shown that upon approaching the transition between 
the RVB phase and the ordered phase, two soft modes develop in the ordered phase, while
two Bragg peaks continuously disappear in the RVB phase.
The continuous behavior of the dimer gap and of the Bragg peaks at the transition point
strongly suggests the second order of the phase transition, consistent with the
theory of Moessner and Sondhi.~\cite{moessner2} Furthermore, our analysis provides 
a more accurate location of the transition point $V/t=0.82 \pm 0.03$ than in previous studies.

A peculiar feature  of the crystal phase is the apparent absence of a singularity
(Bragg peak or soft mode) at the zone corner. We interpret this effect as a strong
reduction of this Fourier harmonics in the crystal without much short-length structure 
in the large 12-site crystal cell. In other words, the $\sqrt{12} \times \sqrt{12}$
crystal may be better approximated by 12-site hexagons without internal structure,
and not by resonating 4-site plaquettes.

It is interesting to compare these new findings with the phase diagram of the same model 
on the square lattice. In that case, the Rokhsar-Kivelson state realized at the 
Rokhsar-Kivelson point ($V/t=1$) is immediately replaced by a plaquette phase, until 
columnar dimer order is finally stabilized for small enough repulsion. The plaquette state 
is a local RVB state on a square plaquette, and in that sense, it is similar to the local 
RVB state on the 12-site diamond. But the analogy can be extended further. The 4-site
plaquette is the smallest subunit that allows resonance between pairs of dimers and
that keep the fundamental point group symmetry of the square lattice, the $C_4$ axis.
The same requirements naturally lead to the 12-site diamond cluster for the triangular
lattice: A 4-site plaquette does not have $C_3$ symmetry, and the 6-site plaquette that
has the shape of a large triangle does not allow pairs of dimers facing each other. So one
can conjecture that there is a general tendency for such models, when the dimer repulsion
is no longer strong enough to stabilize the fully resonating state,
to develop local RVB order in symmetric unit cells that allow sufficient resonance. 

Finally, although we know quite a lot by now about the QDM on the triangular lattice,
a number of very interesting questions remain, among which the role of temperature, the 
nature of the excitation spectrum in the RVB phase, and the possible role 
of visons at the transition to the ordered phase. 

\begin{acknowledgments}
We thank S. Korshunov for discussions on some symmetry related
aspects of the problem. This work was supported by the Swiss National Fund and
by MaNEP. F.B. and M.F. are partially supported by CNR-INFM and by MIUR (COFIN-2004).
F.B. thanks EPFL for the kind hospitality during the accomplishment of this work.
\end{acknowledgments}

\end{document}